\documentclass[prc,twoside,superscriptaddress,showkeys,showpacs,amsmath,
 twocolumn]{revtex4-1}

\usepackage{graphicx, latexsym, amssymb, amsmath, color, multirow, mathrsfs,
CJK,ifpdf,xspace,braket}
\usepackage[section]{placeins}
\usepackage{amsmath}
\usepackage{amsfonts}
\usepackage{amssymb}
\usepackage{bm}% bold math
\usepackage{graphicx}
\usepackage{color} 
\usepackage{txfonts}
\usepackage{hyperref}

\begin{document}

\preprint{APS/123-QED}

\title{Quenching of nuclear matrix elements for $0\nu\beta\beta$ decay by chiral
two-body currents}
\author{Long-Jun Wang}
\affiliation{Department of Physics and Astronomy, University of North Carolina, Chapel Hill, North Carolina, 27599-3255, USA}
\author{Jonathan Engel} \email{engelj@physics.unc.edu}
\affiliation{Department of Physics and Astronomy, University of North Carolina, Chapel Hill, North Carolina, 27599-3255, USA}
\author{Jiang Ming Yao}
\affiliation{Department of Physics and Astronomy, University of North Carolina, Chapel Hill, North Carolina, 27599-3255, USA}
\affiliation{FRIB/NSCL Laboratory, Michigan State University, East Lansing,
Michigan, 48824, USA}

\date{\today}
%%%%%%%%%%%%%%%%%%%%%%%%%%%%%%%%%%%%%%%%%%%%%%%%%%%%%%%%%%%%%%%%%%%%%%%%%%%%%%%%%%%%%%%%%%%%%%%%%%%%%%%%%%%%%%%%%%%%%%%%%%%%%%%%%%%%%%%%%%%%%%%%
\begin{abstract}
We examine the leading effects of two-body weak currents from chiral effective
field theory on the matrix elements governing neutrinoless double-beta decay.
In the closure approximation these effects are generated by the product of a
one-body current with a two-body current, yielding both two- and three-body
operators.  When the three-body operators are considered without approximation,
they quench matrix elements by about 10\%, less than suggested by prior work,
which neglected portions of the operators.  The two-body operators, when treated
in the standard way, can produce much larger quenching.  In a consistent
effective field theory, however, these large effects become divergent and must
be renormalized by a contact operator, the coefficient of which we cannot
determine at present.  
\end{abstract}

\pacs{23.40.-s, 12.39.Fe, 21.60.Cs, 23.40.Hc} % PACS, the Physics and Astronomy

\maketitle

%\section{INTRODUCTION}\label{intro}

Neutrinoless double-beta ($0\nu\beta\beta$) decay is a still hypothetical
process in which two neutrons decay to two protons and two electrons, without
emitting neutrinos \cite{Jon2008Review}.  Its discovery would show that
neutrinos are their own antiparticles and could both pin down uncertain neutrino
masses and discover entirely new new particles.  Experiments to observe the
decay are thus growing in size and cost. Interpreting them, however, requires us
to know the values of the nuclear-matrix elements that figure in the decay rate.
These cannot be measured, only calculated, and theorists have worked
increasingly hard to compute them accurately
\cite{Vergados2012RPP,Jon2017Review}.

Because $0\nu\beta\beta$ decay has never been observed, one really ought to
calculate its matrix elements from first principles, with ingredients that allow
an error estimate.  The standard scheme for doing this is chiral effective field
theory (EFT) \cite{ChiralEFT2009}.  Roughly speaking, one writes down all
interactions among nucleons and pions that are consistent with
spontaneously-broken chiral symmetry.  There are infinitely many of these but a
power-counting scheme in nuclear momenta or the pion mass (both denoted by $Q$)
divided by a QCD scale $\Lambda$ near a GeV allows one to fit all the terms
necessary to achieve any desired level of accuracy, at least in principle.  The
counting is not rigorous, but usually works well.

The weak nuclear current can also be represented in this way.  The leading piece
involves the usual Gamow-Teller and Fermi operators associated with a single
nucleon.  Three orders down in the counting, two-body current operators appear
\cite{Park2003}.  Two-body axial weak currents are currently receiving a lot of
attention because they appear \cite{quenching2018} to mostly explain the
longstanding tendency of nuclear theorists to over-predict single-$\beta$ decay
rates \cite{Brown1988review,quenching2015}, forcing them to adopt an effective
value for the axial-vector coupling constant $g_A$ that is significantly smaller
than the bare value.  Recent suggestions \cite{Iachello2015} that $g_A$ should
exhibit similar quenching in $0\nu\beta\beta$ matrix elements, where it is
squared and would thus have a larger impact, have led theorists to examine the
effects of two-body current operators in $0\nu\beta\beta$ decay.  Ref.\
\cite{Javier2011PRL} was the first work on the issue.  The authors, and those of
the later QRPA-based work of Ref.\ \cite{Jon2014PRC}, normal-ordered the
two-body operators with respect to the non-interacting ground state of spin- and
isospin-symmetric nuclear matter to obtain an effective density-dependent
one-body current that quenched $0\nu\beta\beta$ matrix elements by roughly 30\%,
less than one might fear because the quenching was less effective when the
virtual neutrino exchanged in the process transferred a significant amount of
momentum from one nucleon to the other.  The assumptions underlying the
conclusions --- that an effective one-body operator is sufficient and that
normal-ordering with respect to a simple nuclear-matter state is sufficient to
obtain it --- have never been examined, however.

Here we carry out a more comprehensive analysis.  We construct the explicit
product of the one-body and two-body current operators, the leading contribution
from two-body currents to the $0\nu\beta\beta$ matrix element in the closure
approximation (which in tests is accurate to 10\% or so
\cite{Muto1994NPA,Senkov2014PRC}), to obtain two- and three-body
$0\nu\beta\beta$ operators.  After an illustrative calculation in symmetric
nuclear matter, we evaluate the matrix elements of these operators between
reasonable approximations to full shell-model wave functions in $^{76}$Ge and
$^{76}$Se, and in $^{48}$Ca and $^{48}$Ti.  (The first pair has been used in
many experiments; see, e.g., Ref.\ \cite{LEGEND2017}.)  We find that the obvious
sources of quenching, involving three nucleons (only two of which decay), have
even smaller effects than the effective-operator approach suggests.
Contributions from pairs of nucleons that both generate the two-body current and
decay themselves turn out to be more problematic, however.

In $0\nu\beta\beta$ decay the weak current acts twice.  The nuclear matrix
element that governs the decay is given by 
\begin{align} 
\label{eq:NM}
M = \frac{4\pi R}{g^2_A} \int \frac{d\bm x_1 d\bm x_2 d\bm q}{(2\pi)^3} \,
\frac{e^{i\bm q \cdot (\bm x_1 - \bm x_2)} }{q(q+E_d)} \, \bra{0^+_F} \hat{\mathcal
J}^\mu(\bm x_1)  \hat{\mathcal J}_\mu(\bm x_2) \ket{0^+_I}\,, 
\end{align}
where $\hat{\mathcal{J}}(\bm x)$ is the nuclear current, $R\equiv1.2A^{1/3}$ fm
is the nuclear radius, $g_A\approx 1.27$, $\bm q$ labels the momentum transfer
and $E_d \equiv \bar E-(E_I+E_F)/2$ is an average excitation energy, to which
the matrix element is not sensitive ($\bar E$ is an absolute average energy).
Up to third order in $Q/\Lambda$, the nuclear current $\hat{\mathcal J}^\mu$ can
be written as $\hat{\mathcal J}^\mu = \hat{\mathcal J}^\mu_{\text{1b}} +
\hat{\mathcal J}^\mu_{\text{2b}}$, where the two terms in the sum are the one
and two-body pieces of the current.  The first of these is \cite{Park2003},
\cite{Simkovic1999PRC, Simkovic2008PRC}
\begin{align} 
\label{eq:1b-current}
\hat{\mathcal J}^\mu_{\text{1b}}(\bm x) = \sum^A_{n=1} \Big[ \delta_{\mu0}
J_{n,0}(q^2) - \delta_{\mu j} \bm J_{n,j}(q^2) \Big] \tau^-_n \delta(\bm x - \bm
r_n) \,.
\end{align}
Here $\bm r_n$ is the coordinate of the $n$th nucleon, $\bm q \equiv
i\bm{\nabla}$ and
\begin{align}
\label{eq:one-body-a}
J_{n,0}(q^2) =&\ g_V + \dots, \\
\label{eq:one-body-b}
\bm J_{n}(q^2) =&g_A\bm\sigma_n + \ i (g_M+g_V)  \frac{\bm\sigma_n\times\bm q}{2m_N} - g_P(q^2) \frac{\bm q \, \bm\sigma_n \cdot \bm q}{2m_N} 
+ \dots \,,
\end{align}
where $g_V=1, g_M\approx 3.706$, $g_P(q^2)$ is given, e.g., in Ref.\
\cite{Javier2011PRL}, and $m_N$ is the nucleon mass.  In what follows, we will
be looking at the axial current, and so neglect contributions of $J_{n,0}(q^2)$.
The terms indicated by ellipses can be shown \cite{Javier2011PRL} to contribute
negligibly to the matrix element in Eq.\ \eqref{eq:NM}.

\begin{figure}[b]
\centering
\includegraphics[width=0.43\textwidth]{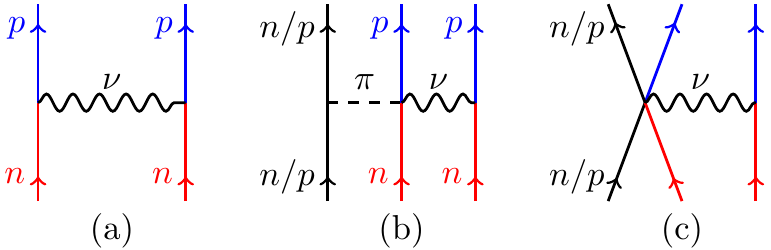}
\caption{\label{fig:diagram} (Color online.) $0\nu\beta\beta$ decay, with
electron lines omitted.  Digaram (a) shows the leading contribution, in which
the one-body current acts twice, turning two neutrons into two protons via the
exchange of a Majorana neutrino.  Diagram (b) shows the action of the
pion-exchange two-body current at one vertex; the line on the left represents
either a proton or a neutron. In diagram (c) the contact current replaces the
pion-exchange current.}
\end{figure}

In considering the two-body current, we neglect the term with coefficient $c_6$
\cite{Park2003} and terms with two-body pion poles \cite{Klos2014PRD}, but
otherwise keep the full momentum-dependence of Ref.\ \cite{Park2003}), Fourier
transforming Eqs.\ (A5) and(A6) of that paper with, following Ref.\
\cite{Krebs2017}, an additional factor of $-1/4$ in the contact term gives the
leading space piece of the axial two-body current operator in coordinate space:
\begin{widetext}
\begin{align} \label{eq:2b-current1}
  \hat{\bm{\mathcal J}}_{\text{2b}}(\bm x) &= \sum^A_{k<l} \bm J_{kl}(\bm x) \,,
\\
\label{eq:2b-current2}
  \bm J_{kl}(\bm x) =&\ 
  \frac{2c_3 g_A}{m_N F^2_\pi} \Big[m^2_\pi \Big( \big( \frac{\bm\sigma_l}{3} -
  \bm\sigma_l \cdot\hat{\bm r}\hat{\bm r} \big)Y_2(r) - \frac{\bm\sigma_l}{3}
  Y_0(r) \Big) + \frac{\bm\sigma_l}{3}\delta(\bm r)\Big] \tau^-_l \delta(\bm x - \bm r_k) 
  \ + \ (k \leftrightarrow l) \nonumber \\ %---c3
  &\ + \big(c_4 + \frac{1}{4}\big) \frac{g_A}{2m_N F^2_\pi}
  \Big[ m^2_\pi \Big( \big( \frac{\bm\sigma_\times}{3} -
         \bm\sigma_k \times \hat{\bm r} \bm\sigma_l \cdot \hat{\bm r}  \big)
         Y_2(r) - \frac{\bm\sigma_\times}{3} Y_0(r) \Big) 
         + \frac{\bm\sigma_\times}{3}\delta(\bm r) \Big] 
         \tau^-_\times \delta(\bm x - \bm r_k)   \ + \ (k \leftrightarrow l)   \nonumber \\ %---c4
         &\ - \frac{g_A}{4m_N F^2_\pi} 
        \Big[ 2\hat d_1 (\bm\sigma_k \tau^-_k + \bm\sigma_l \tau^-_l) + \hat d_2 \bm\sigma_\times \tau^-_\times \Big] \delta(\bm r)
        \delta(\bm x - \bm r_k ) \,, 
\end{align}
\end{widetext}
where $F_\pi = 92.4$ MeV is the pion decay constant, $m_\pi$ is the pion mass,
$\bm r = \bm r_k - \bm r_l$ and $\hat{\bm r} \equiv \frac{\bm r}{r}$.  The
Yukawa functions $Y$ are $Y_0(r) = \frac{e^{-m_\pi r}}{4\pi r}$ and $Y_2(r) =
\frac{1}{m^2_\pi} r \frac{\partial}{\partial r} \frac{1}{r}
\frac{\partial}{\partial r} Y_0(r)$, and the compound spin and isospin operators
are $\bm\sigma_\times = \bm\sigma_k \times \bm\sigma_l$ and $\tau^-_\times =
(\tau_k \times \tau_l)^-$ \cite{Park2003}. 
The product of currents in Eq.\ \eqref{eq:NM} for the $0\nu\beta\beta$ matrix
element can be broken up into contributions from one- and two-body currents.
The leading piece, from two one-body currents acting as in diagram (a) of Fig.\
\ref{fig:diagram}, is what has been considered almost exclusively in prior work.
The first correction comes from diagrams like (b) and (c), in which one of the
one-body currents is replaced by a two-body current, of either long range
(diagram (b) with an internal pion) or short range (diagram (c)).  Ref.\
\cite{Javier2011PRL} first considered these contributions, but only
approximately, as we've already mentioned; here we consider them more
completely.  

%-------The Feynman diagram here:
%
%
%-------Goldstone diagrams figure:
\begin{figure}[b]
\centering
\includegraphics[width=0.5\textwidth]{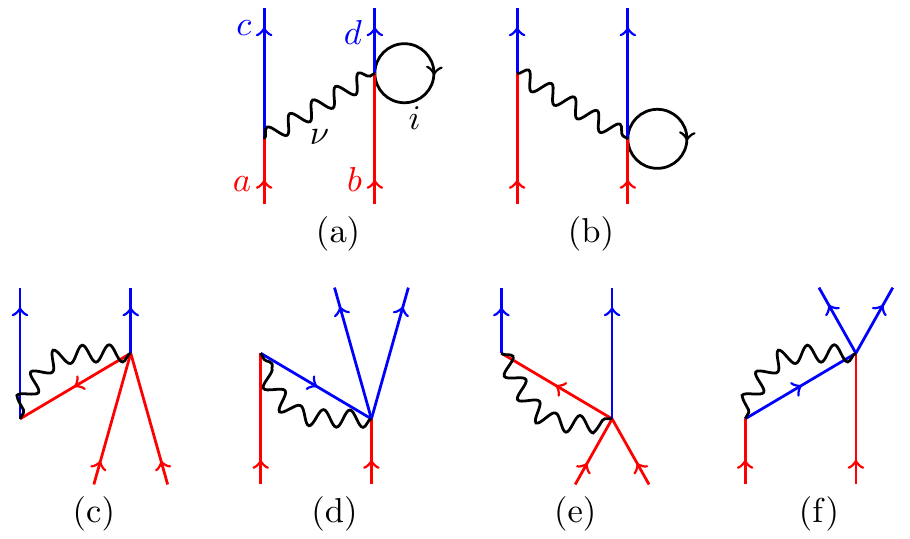}
\caption{\label{fig:fermi-diags} (Color online.) Contributions to the
$0\nu\beta\beta$ matrix element in symmetric nuclear matter.  Red lines
represent neutrons, blue lines protons, and wiggly black lines the exchanged
neutrino.  The top row of diagrams (a) and (b) represent the contributions
considered in Ref.\ \cite{Javier2011PRL}.  The diagrams in the bottom row (c--f)
have not been considered before.}
\end{figure}

To get an idea of what to expect in real nuclei, we begin with a more schematic
discussion of nuclear matter, modeled after that in Ref.\ \cite{Javier2011PRL}.
To simplify matters here (and only here), we neglect all but the $d_1$ and $d_2$
contact pieces of the two-body current (see Eq.\ \eqref{eq:2b-current2}) and
evaluate all the current operators at $q=0$.

In nuclear matter, the one-body-two-body contributions just alluded to can be
represented by the Goldstone-Heugenholtz diagrams in Fig. \ref{fig:fermi-diags}.
The top row of diagrams, in which one nucleon in the two-body current is a
spectator, was treated in Ref.\ \cite{Javier2011PRL}.  The spectators contribute
coherently, leading to a factor of the nuclear density in the matrix element,
and allowing one to replace the two-body current in the diagram by a
density-dependent one-body effective current.  Three-body operators need never
be considered explicitly in such a procedure. 
  
The bottom row has not been examined before.  These diagrams involve the
contraction of creation and annihilation operators from different vertices and
superficially are perhaps not as coherent.  But the internal hole and particle
lines are summed and it is not obvious that the contributions of these diagrams
will be much smaller.  It is obvious, however, that diagrams (e) and (f) will
have the same sign as the top row of diagrams, and that diagrams (c) and (d)
will have the opposite sign.  A diagram's sign contains a factor of
$S=(-1)^{n_h+n_l}$, where $n_h$ is the number of hole lines and $n_l$ is the
number of nucleon loops.  The diagrams in the top row have one hole line and one
nucleon loop, and thus $S=1$.  Diagrams (e) and (f) have no hole lines and no
nucleon loops ($S=1$) and diagrams (c) and (d) have one hole line and no nucleon
loops ($S=-1$).  The net effect once all terms are taken into account remains to
be seen.  

We evaluate the diagrams in the closure approximation, that is, by neglecting
the variation in the energies of the intermediate particles and holes in the
bottom row of diagrams.  To simplify matters, we set $E_d$ in Eq.\ \eqref{eq:NM}
to zero, so that the energy denominators contain just the $1/q^2$ associated
with the neutrino. We take the external momenta $\bm k_a$, $\bm k_b$, $\bm k_c$,
and $\bm k_d$, which are to represent those of valence nucleons, to lie on the
Fermi surface ($k=k_F$), though in evaluating the angle average of
$1/|k_a-k_c|^2$ in the top row of diagrams we let the magnitude of one of the
two momenta be distributed with equal probability in a symmetric interval of
width $k_F$ around the Fermi surface (to avoid a divergent result).  With these
assumptions, the amplitude represented by each of the diagrams has the form 
\begin{equation}
\label{eq:diag-form}
X \,
\delta(\bm{k}_a + \bm{k}_b - \bm{k}_c -\bm{k}_d)
\bra{f}\bm{\sigma}_1 \cdot \bm{\sigma}_2 \, \tau_1^- \tau_2^- \ket{i}\,, 
\end{equation}
for some constant $X$, where the matrix element refers just to the spin-isospin
part of the initial ($i)$ and final ($f$) wave functions.  We separately sum
diagrams $(a)$ and $(b)$, $(c)$ and $(d)$, and $(e)$ and $(f)$ (the members of
each pair are equal).  The results:
\begin{equation}
\label{eq:X}
\begin{aligned}
X_{ab} &\equiv X_\text{(a)}+X_\text{(b)} \approx   -\frac{2C(2+2\ln2)k_F}{3\pi^2} \\
X_{cd} &\equiv X_\text{(c)}+X_\text{(d)} = \frac{3 C k_F}{4 \pi^2} 
\approx - \frac{1}{2} X_{ab} \\
X_{ef} &\equiv X_\text{(e)}+X_\text{(f)} \approx -\frac{6C(\Lambda-k_F)}{4\pi^2}
% \approx -4 X_{cd}
\approx 2 X_{ab} \,,
\end{aligned}
\end{equation}  
where $C$ is a constant containing $d_1$, $d_2$, the nuclear radius, $F_\pi$,
$g_A$, and the nucleon mass, and where we take $\Lambda$, the momentum at which
we cut off the integral over particle states, to be $3k_F$.  The contribution of
diagrams (e) and (f) would be reduced by avoiding the closure approximation (the
energy denominator would increase by an amount that would reach about 40\% by
the upper limit of the integral) but would still grow with $\Lambda$.  The
relative signs of the contributions reflect the discussion above. 

We can break the results of Eq.\ \eqref{eq:X} into contributions of three-body
operators, with $n\ne k,l$ in the products of the currents in Eqs.\
\eqref{eq:NM}, \eqref{eq:1b-current}, and \eqref{eq:2b-current1}, and two-body
operators, with $n=k$ or $l$. In addition to the producing the quenching
contributions $X_{ab}$ discussed in Ref.\ \cite{Javier2011PRL}, three-body
operators also contribute exactly twice $X_{cd}$, so that the net quenching
produced by the three-body operators nearly vanishes.  Two-body operators
produce $X_{ef}-X_{cd}$, which is about 5/2 $X_{ab}$ (a number, that, again,
would be a bit smaller without closure) so the final overall quenching is
greater than obtained in prior work.  As we see next, conclusions much like
these still obtain when we use realistic nuclear wave functions, nucleon form
factors, and the full two-body current. 

One might argue that in computing $X_{ef}$ we should not use a cutoff to
regulate the integral.  In a more consistent chiral effective field theory like
that in Refs.\ \cite{Cirigliano2017,Cirigliano2018}, in which all two-body
processes such as those in diagrams (e) and (f) are evaluated in isolation and
the results subsequently embedded in a many-body calculation (so that Eq.\
\eqref{eq:NM} is not the starting point), that is standard practice; dimensional
regularization restricts the momenta in loops to be low.  But that procedure
introduces counter terms with unknown coefficients at chiral orders below those
considered here. We are simply trying to assess the quenching induced by
two-body currents alone, and a cutoff simulates the effects of nucleon form
factors in the sum over intermediate states in a realistic calculation. Of
course, the use of form factors in conjunction with chiral currents is not
consistent; if we really want to do EFT we will require counter terms.  We
return to this issue later.

First, however, we present realistic shell-model-like calculations of the decay
matrix elements for $^{48}$Ca and $^{76}$Ge.  We carry these out in typical
oscillator valence spaces: the $fp$ shell for the lighter nucleus (and the final
nucleus $^{48}$Ti) and the $f_{5/2}pg_{9/2}$ space for the heavier one (and for
$^{76}$Se).  Here, without the ability to include a complete set of
intermediate-nucleus states, we need to work to evaluate the matrix elements of
three-body operators.  We do so by combining the three-body matrix elements of
the operator $\hat{O}_\text{3b}$ (representing the three-body part of
$\hat{\mathcal J}^\mu(\bm x_1)_{\text{1b}} \hat{\mathcal J}_\mu(\bm
x_2)_{\text{2b}} + \hat{\mathcal J}^\mu(\bm x_1)_{\text{2b}} \hat{\mathcal
J}_\mu(\bm x_2)_{\text{1b}}$) with three-body transition-densities to obtain
\begin{equation} 
\label{eq:3body}
  M^{\text{3b}} 
  = - \sum_{abcdef}
  \bra{abc} \hat O_{\text{3b}} \ket{def}
  \rho^{\text{3b}}_{abc,def} \,,
\end{equation}
where
\begin{equation}
\label{eq:tdens}
\rho^\text{3b}_{abc,def} = \bra{0^+_F}a^\dag_a a^\dag_b a^\dag_c a_d a_e a_f
\ket{0^+_I} \,. 
\end{equation}
Here the subscripts $a,b,\dots$ represent full single-particle labels, e.g., $a$
stands for the set $\{\tau_a,n_a,l_a,j_a,m_a\}$, i.e.\ the isospin, harmonic
oscillator radial quantum number, orbital angular momentum, total angular
momentum, and $z$-projection associated with the level in question.
$M^\text{3b}$ is thus the three-body piece of the matrix element $M$ in Eq.\
\eqref{eq:NM}.

We calculate the three-body matrix elements of $\hat O_{\text{3b}}$ in much the
same way as the matrix elements of three-body interactions were calculated in
the work of Refs.\ \cite{Petr2006PRC, Petr2007FBS, Roth2014PRC}, i.e., we first
compute them in a large three-body Jacobi basis and then transform to a coupled
product basis.  To obtain $\rho^{\text{3b}}$ we use the generator coordinate
method (GCM) to approximate shell-model wave functions \cite{Thomas2010PRL}.  As
in Ref.\ \cite{Jiao2017PRC}, we use the Hamiltonian KB3G \cite{Poves2001NPA} for
the nuclei with 48 nucleons and the Hamiltonian GCN2850 \cite{Javier2009NPA} for
those with 76, and include both axial deformation and isoscalar pairing
\cite{Nobuo2014} as generator coordinates.  We assume that the valence space
sits on top of an inert core of filled oscillator levels.  If all three nucleons
acted on by $\hat{O}_\text{3b}$ are in the valence space, the densities
$\rho^\text{3b}$ are the matrix elements between the initial and final GCM
states of three creation and three annihilation operators.  If one of the three
nucleons comes from the shell model core, on the other hand, then the
$\rho^{\text{3b}}$ reduce to simpler two-body valence-space transition
densities.  The corresponding contributions to $M^\text{3b}$ are what one would
obtain by normal ordering the product of currents with respect to the inert
shell-model core, a more realistic version of the symmetric nuclear-matter state
considered in Ref.\ \cite{Javier2011PRL}.  The contractions generated by the
normal-ordering can be either between creation and annihilation operators within
the two-body current, as in the top row of Fig.\ \ref{fig:fermi-diags} and in
Ref.\ \cite{Javier2011PRL}, or between operators from different currents, as in
the bottom row of Fig.\ \ref{fig:fermi-diags}.  

\begin{figure}[t]
  \includegraphics[width=0.5\textwidth]{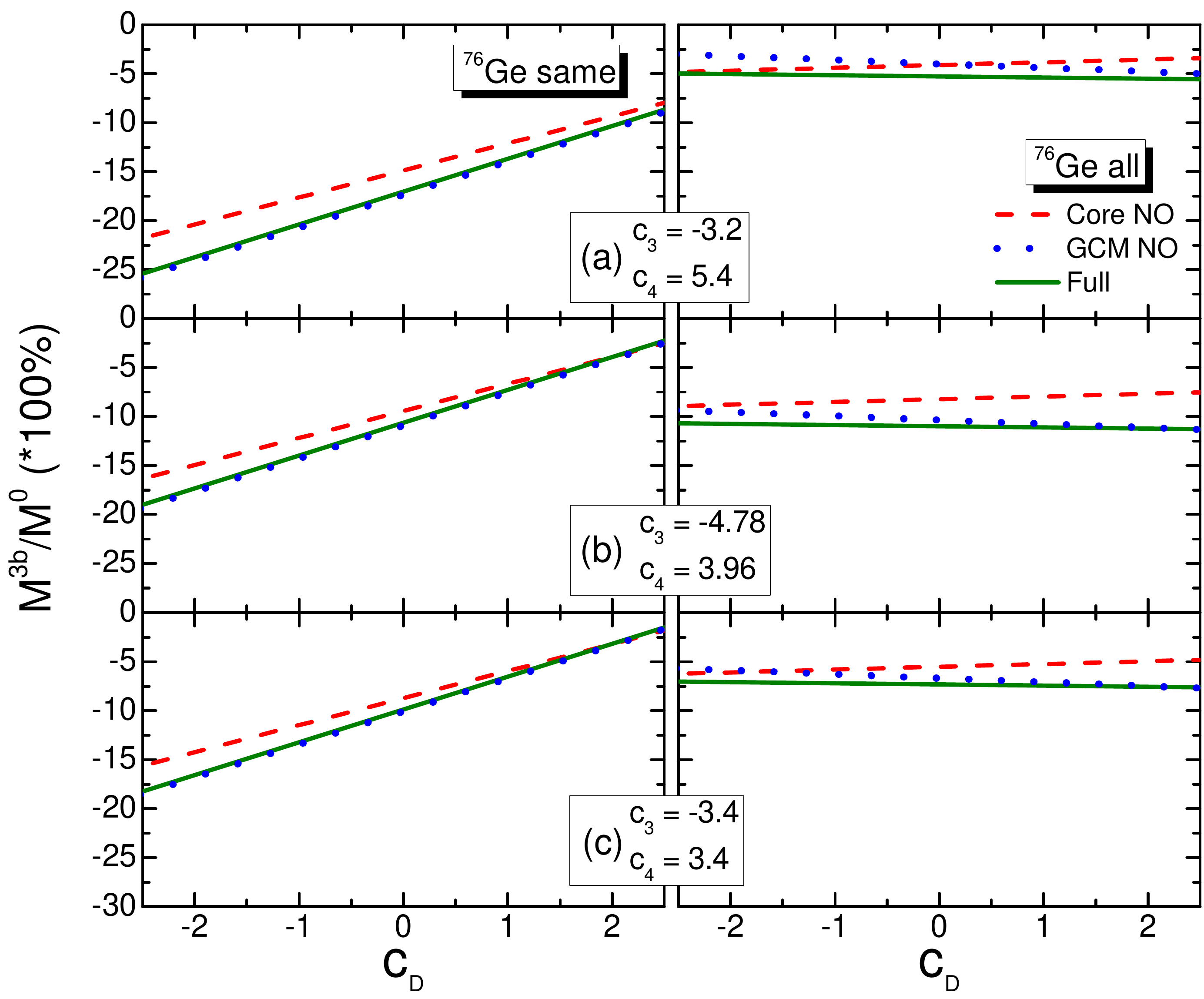} 
  \caption{\label{fig:Ge76-3b} (Color online.) Relative effects on the
  $0\nu\beta\beta$ matrix element from the three-body-operator parts of diagrams
  involving chiral two-body currents (as shown in Fig.\ \ref{fig:diagram}(b) and
  (c), and Eq. (\ref{eq:3body}), with several sets of coefficients $c_3, c_4$,
  and as a function of $c_D$ for $^{76}$Ge. The solid line represents the full
  results, the dashed line the approximate results when three-body operator are
  discarded after normal-ordering with respect to the inter core, and the dotted
  line the results when the normal-ordering is with respect to an ensemble
  containing the GCM $^{76}$Ge and $^{76}$Se ground states.  The results in the
  panels on the left include only contributions from the contraction of creation
  an annihilation operators at the same vertex in Fig.\ \ref{fig:diagram}.  See
  text for details.  }
\end{figure}

\begin{figure}[t]
  \includegraphics[width=0.49\textwidth]{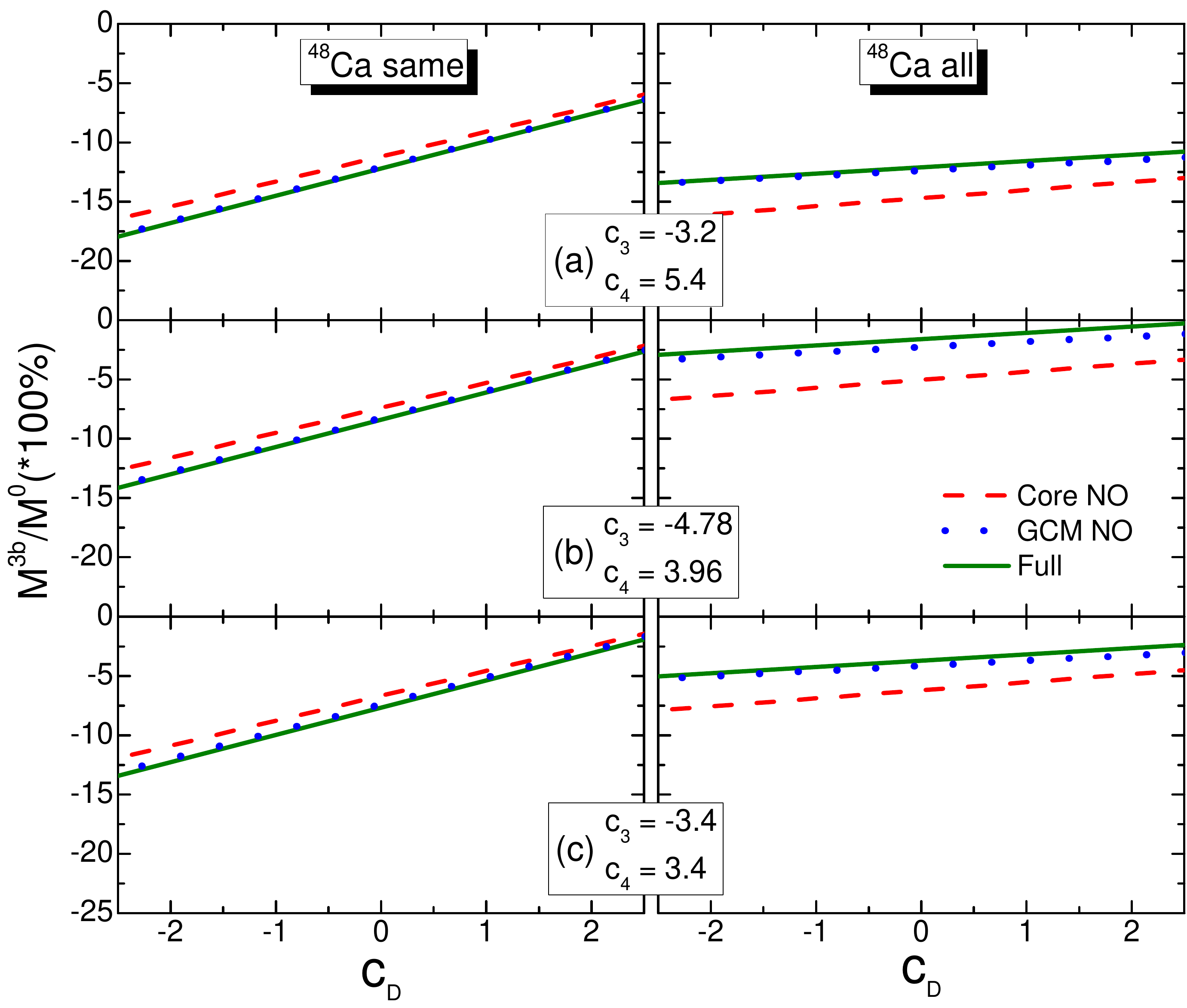} 
  \caption{\label{fig:Ca48-3b} (Color online.) Same as Fig.\ \ref{fig:Ge76-3b}
  but for $^{48}$Ca.}
\end{figure}

Figure \ref{fig:Ge76-3b} shows the ratio $M^{3b}/M^0$, where $M^0$ is the
leading part of the matrix element that comes from one-body currents at both
vertices (Fig.\ \ref{fig:diagram}(a)) for the decay of $^{76}$Ge, with the GCM
wave functions described in the previous paragraph.  These wave functions are
not quite as complex as those in Ref.\ \cite{Jiao2017PRC}; they are linear
combinations of states with a single value for the isoscalar pairing amplitude
and seven values for the axial deformation parameter $\beta$.  The resulting
matrix element --- 3.47 --- is reasonably close to the exact result of 2.81
\cite{Javier2009NPA}.  The different panels in the figure correspond to
different values for the couplings $c_3$, and $c_4$, and we present them as
functions of $c_D \equiv d_1+2d_2$.  The values $c_3=-3.2, c_4=5.4$ are from
Ref.\ \cite{EM}, $c_3=-4.78, c_4=3.96$ from Ref.\ \cite{PWA}, and $c_3=-3.4,
c_4=3.4$ from Ref.\ \cite{EGM}.  To get the results on left side of the figure
(labeled ``same''), we include only the contributions of contractions of
creation and annihilation operators from within the same (two-body) current,
like those of Ref.\ \cite{Javier2011PRL} or diagrams (a) and (b) in Fig.\
\ref{fig:fermi-diags}.  (Note, however, that Ref.\ \cite{Javier2011PRL} omitted
the factor of $-1/4$ in the last line of Eq.\ \eqref{eq:2b-current2}.) We
include all possible contractions to obtain the results on the right.  The
dashed and dotted lines show approximate results in which the we have discarded
three-body operators that survive normal ordering with respect to the inert core
(the discarded terms are those in which all three nucleons are in the valence
shell) and with respect to an ensemble containing the full GCM ground states of
$^{76}$Ge and $^{76}$Se, weighted equally.  The ideas on which this ensemble
normal ordering is based are presented in Ref.\ \cite{Stroberg2017PRL}.

The figure shows that with only the contractions from within the two-body
current, the three-body operators quench the matrix element by $5\%$ to $25\%$
for $|c_D| \le 2$. This level of quenching is what one would obtain with the
density-dependent effective-operator treatment of Ref.\ \cite{Javier2011PRL} at
a somewhat lower nuclear density than that taken in Ref.\ \cite{Javier2011PRL}.
A similar level of quenching holds in single-$\beta$ decay, as discussed in Ref.
\cite{quenching2018}.  

When all the contractions are included, the quenching decreases, just as in our
nuclear-matter results for the contact part of the current. In the bottom two
panels it doesn't decrease very much, but in the top panel it decreases
significantly.  The full results are also nearly independent of $c_D$; the
almost complete cancellation between the different three-body contractions we
found in the nuclear-matter calculation is borne out here.  When all is said and
done, the three-body operators end up quenching matrix element by 5 or 10\%.

A final observation about Fig.\ \ref{fig:Ge76-3b}: the normal ordering with
respect to the inert core indeed provides most of the matrix element, with the
configurations in which all three nucleons are in the valence shell contributing
relatively little.  The normal ordering usually gets even better when we do it
with respect to the more realistic reference ensemble.  That is good news for
many-body calculations in which three-body operators are problematic.

Figure \ref{fig:Ca48-3b} shows the all the same results for $^{48}$Ca.  The
overall effects of two-body currents are comparable, though the two interactions
for which the new contractions do little to the $c_3$ and $c_4$ parts of the
current in $^{76}$Ge do more here (and vice versa).

\begin{figure}[t]
  \includegraphics[width=0.40\textwidth]{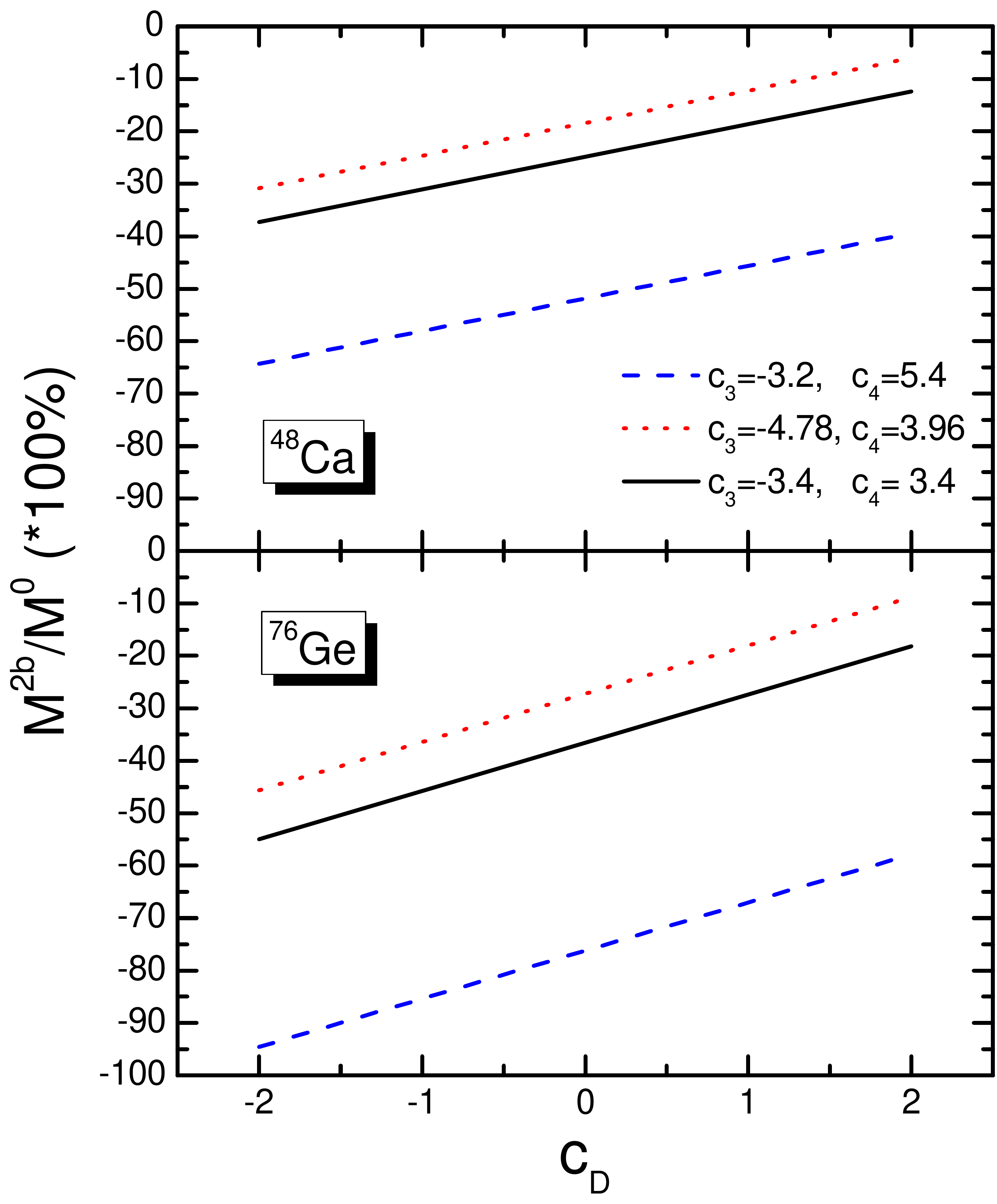} 
  \caption{\label{fig:Ge76-2b} (Color online.) Relative effects on the
  $0\nu\beta\beta$ matrix element from the two-body-operator parts of diagrams
  involving chiral two-body currents (as shown in Fig.\ \ref{fig:diagram} (b)
  and (c), with several sets of coefficients $c_3, c_4$, and as a function of
  $c_D$, for $^{48}$Ca (top), and $^{76}$Ge (bottom).}
\end{figure}

We turn finally to the troublesome two-body operators in the product of one-body
and two-body currents.  These operators have their origins in loops, and as
already noted, without nucleon form factors and/or other regulators they produce
divergences.  The operator that comes from the contact current, for example, is 
\begin{align} \label{eq:2b-cD}
\hat{O}&^\text{2b}_{c_D} = \\ 
&\frac{2c_D R}{\pi m_N F^2_\pi} \sum^A_{k \ne l} 
\int d q \frac{\big[q g^2_A(q^2) - \frac{q^3
g_A(q^2)g_P(q^2)}{6m_N}\big]}{(q+E_d) g^2_A}  
\bm\sigma_k \cdot \bm\sigma_l \tau^-_k  \tau^-_l \delta(\bm r) \,. 
\nonumber
\end{align}
The integral diverges if $g_A$ has no $q$ dependence.  Here, for the purposes of
estimation, we assign it the dipole nucleon form factors given in Ref.\
\cite{Simkovic2008PRC} and used in nearly every prior calculation.  Figure
\ref{fig:Ge76-2b} shows the relative effects on the nuclear matrix elements from
all the two-body operators and with the GCM wave functions described earlier.
These operators can quench the matrix element substantially, by amounts that
range up to more than 90\% in $^{76}$Ge and 60\% in $^{48}$Ca.  The amount of
quenching, however, is very sensitive to $c_3, c_4$ and $c_D$, and can be as
small as $10\%$ in $^{76}$Ge.  

All these results could be changed somewhat by the terms we've omitted, which
include three-body tensor pieces, pion poles in the two-body current, operators
that come from the action of two two-body currents (one at each vertex), and
even higher-order currents in chiral EFT.  The effects of the last two are
nominally smaller, but could be significant because of cancellations between
lower-order contributions. And again, the closure approximation exaggerates the
quenching somewhat.  All of this is secondary, however, to the meaning of the
large and parameter-dependent quenching by two-body operators.  Their
contributions to the matrix element reflect the scale $\Lambda$ associated with
the nucleon form factors, which are not consistent with chiral EFT.  Within
effective field theory, we instead require a contact counter term to renormalize
the loop diagrams that produce the two-body operators.  The coefficient of that
term is unknown, and there is no obvious way to use data to fix it.  

The necessary counter term is in fact already a part of the analysis in Ref.\
\cite{Cirigliano2017}, where it occurs one order below that of the two-body
currents.  With our form-factor regulator, which simulates a cutoff, that is the
order required to cancel the divergent loops.  In the dimensional-regularization
scheme of that paper, however, the two-body operators, after removal of the
divergence, would naturally contribute at the same order as the three-body
operators.  (We might even expect their contribution to be bit smaller because a
typical momentum transfer and the pion mass are both less than $k_F$.)  But then
another counter term at this same order, with an equally unknown coefficient,
would have to be included as well.

In the end, there are two options.  One can work with effective field theory
consistently from the beginning, in which case our results with realistic wave
functions show that show that quenching from the leading three-body operators in
the product of currents is probably around 10\% (``probably'' because of the
potential effects of pion poles and tensor terms). That amount of quenching is
less than previous work suggests, and nearly independent of $c_D$.  Furthermore,
in the future we can compute the effects of these operators to a good
approximation by discarding all but the normal-ordered two-body pieces.  Within
the EFT framework, however, the nominally similar contributions of two-body
operators must be supplemented by those of contact operators with unknown
coefficients.  And similar contact operators occur at lower orders in
$Q/\Lambda$, including leading order \cite{Cirigliano2018}.  We may have to wait
to get a good handle on the coefficients until lattice QCD is up to the job.  

The alternative is to work in an old-fashioned model, with explicit heavy mesons
and nucleon form factors, leading to potentially substantial quenching from
two-body operators.  Even this framework, however, will lead to additional
short-range contributions to $0\nu\beta\beta$ decay from heavy-meson exchange
currents, and no way of systematically estimating error.  EFT, even without a
rigorous power counting, is probably the better route, and the degree to which
operators beyond the leading chiral order alter matrix elements thus remains to
be seen in full.

%--------Acknowledgement:
We thank P. Gysbers and P. Navr\'atil for helping with the three-body Jacobi
basis, and E.\ Mereghetti for an illuminating calculation in chiral EFT. We also
thank V. Cirigliano, J. D. Holt, C. F. Jiao, J.  Men\'endez, T.\ Papenbrock, and
S.\ Pastore for valuable discussions, and F.\ Nowacki and A.\ Poves for allowing
us to use the unpublished shell-model interaction GCN2850.  Finally, we
acknowledge the hospitality of the Institute for Nuclear Theory at the
University of Washington.  This work was supported by U.S.  Department of Energy
Grant Nos.\  DE-FG0297ER41019 and DE-SC0008641. 

%%%%%%%%%%%%%%%%%%%%%%%%%%%%%%%%%%%%%%%%%%%%%%%%%%%%%%%%%%%%%%%%%%%%%%%%%%%%%%%%%%%%%%%%%%%%%%%%%%%%%%%%%%%%%%%%%%%%%%%%%%%%%%%%%%%%%%%%%%%%%%%

%\bibliographystyle{apsrev}	% (uses file "plain.bst"unsrt,abbrv,alpha,apsrev4-1)
%\bibliography{currents}	% expects file "myrefs.bib

\end{document}